# A Novel Service Oriented Model for Query Identification and Solution Development using Semantic Web and Multi Agent System

Muneendra Ojha
Indian Institute of Information Technology - Allahabad
Deoghat Jhalwa
Allahabad 211012

## ABSTRACT
In this paper, we propose to develop service model architecture by merging multi-agentsystems and semantic web technology. The proposed architecture works in two stages namely, Query Identification and Solution Development. A person referred to as customer will submit the problem details or requirements which will be referred to as a query. Anyone who can provide a service will need to register with the registrar module of the architecture. Services can be anything ranging from expert consultancy in the field of agriculture to academic research, from selling products to manufacturing goods, from medical help to legal issues or even providing logistics. Query submitted by customer is first parsed and then iteratively understood with the help of domain experts and the customer to get a precise set of properties. Query thus identified will be solved again with the help of intelligent agent systems which will search the semantic web for all those who can find or provide a solution. A workable solution workflow is created and then depending on the requirements, using the techniques of negotiation or auctioning, solution is implemented to complete the service for customer. This part is termed as solution development. In this service oriented architecture, we first try to analyze the complex set of user requirements then try to provide best possible solution in an optimized way by combining better information searches through semantic web and better workflow provisioning using multi agent systems.

## 1. INTRODUCTION
Long gone are the days of dumb terminals running set algorithms and producing results which otherwise if done manually would have taken much more time. Computer scientists are aiming for intelligent programs. The ultimate target of artificial intelligenceis to build programs that are capable of performing like or better than humans. Since this cannot be achieved in a day, researchers target towards finding solutions which give better result than the already existing ones. Acting like humans is an extremely complex task so AI focuses more on solving smaller problems which when clubbed together creates the platform for solving or implementing a real life scenario. Agent system is one such example where small programs called agents collectively try to achieve the best possible solution of a complex problem. Ideology of agent based system revolves around creating complex systems as a composition of simpler, heterogeneous, distributed components wherein intelligence comes with ability to observe and learn from environment. A vast domain of problems is being addressed using multi agent research and technology [1], [2], [3], [4], [5].

Although agent based computing has been around for some time but it is only in the last decade that it has found interest of people beyond research community. Typical software architectures containdynamically interacting modules, each with their own thread and engaging in complex coordinationprotocols [6]. Agent based computing presents a novel software engineering paradigm emerging from the amalgamation of object oriented distributed computing and artificial intelligence [7]. In this paper, we advocate an agent-oriented paradigm, conceptualizing the analysis and design of an agent-based system for providing web oriented services. The remainder of this paper is structured as follows. In the second section, we provide motivation for this architecture. In the third section we provide elaborate view of the intelligent agents, how they perceive, interact and perform. We also introduce the idea of manual agents and how do autonomous agents interact with them to optimize the perception. In the fourth section, we provide the agent oriented model for our proposed architecture and describe the agent oriented analysis and design. We also compare it against traditional object oriented design methodologies. Finally, in the last section, we summarize our results by providing an insight into the future challenges.

## 2. MOTIVATION
Most of the real world applications are extremely complex than they were few years hence. They contain components which are dynamically interacting with environment through their own process threads. Most software engineering paradigms are unable to provide structure that could handle such complexities. It is hard to identify a general framework that allows the management of service in a standard platform independent way. So we propose a middleware architecture that functions as an intermediary between consumer and service provider. Although the full interoperability between various middleware platforms is still not achieved, this architecture serves as a valid starting point to bring distributed services together. In order to support services within this service oriented architecture, a lightweight framework is needed that realizes interoperability between different middleware technologies. Both the Model Driven Architecture (MDA) [8] and Service Oriented Architecture (SOA) [9] approaches serve to address the same issue [10]. Let us take an example where a patient needs to see a doctor. Patient is not sure of which specialist she should visit or where she could find a good doctor. Patient uploads the ailment related information over the portal and requests for a good doctor nearby. System parses the information and gathers knowledge out of it. Based on this knowledge, it contacts suitable agents called domain experts, whether autonomous or manual and tries to clearly identify the disease. Once it is identified, it again tries to find out the solution, which in this case would be to find the suitable doctor in a location assumed to be in the patient's vicinity. Then it would contact the doctor, brief him with patient's problem, take an appointment and arrange for a ride if needed by the patient at the set date. Meanwhile,the patient would be able to keep track of the whole process. Patient can also alter the input details as and when required so as to keep system updated about her condition. Idea of the proposed framework is to bring together the vast amount of information or knowledge spread over Internet in a meaningful way so as to serve the specific needs of a consumer. So in our application, SOA better serves the need as it starts from an approach to distributed computing treating software resource as service available over the Internet.

Another example can be of a business person willing to setup a vacuum cleaner factory in some country where it is not a household necessity. Aim is to build a system intelligent as well as powerful enough to understand the requirements of setting up a production unit, technical, financial, social implications, legalities involved, infrastructure details and any other modalities necessary for the queried task. No one can argue against the fact that every related knowledge is available over the Internet and that too free of cost. Question is "how does an agent understand the information and derive the relevant knowledge?" Here lies the actual potential of merging agent oriented approach and Web Service Architecture (WSA) [11][12] in the context of Semantic Web. The Semantic Web provides a common framework that allows data to be shared and reused across application, enterprise, and community boundaries [13]. It enables knowledge sharing among heterogeneous, distributed and dynamic sources. Agents are particularly suitable for performing these kinds of tasks [14]. In the following section, we describe our service architecture and how semantic web enhances its performance. Then, we impress upon the reasons that make multi-agent systems a strikingly appropriate and pertinent solution for the design and development of a service oriented system over the web. In this context, we focus upon the role of Agent Oriented Software Engineering (AOSE) in the Web Service Architecture.

## 2.1 Semantic Web and Web Service Architecture

Semantic Web was conceptualized by the inventor of World Wide Web with a vision to exploit the full potential of data available over Internet. As per Lee:

*"If HTML and the Web made all the online documents look like one huge book, RDF, schema, and inference languages will make all the data in the world look like one huge database"*[16]

Semantic web describes the relation between things on website as well as their properties. Instead of visualizing Internet as a web of pages, it builds a web of data. It starts with defining Uniform Resource Identifier (URI) which is a string of characters used to identify name of a resource on Internet. Resource can be anything ranging from books, cars, people, web pages, digital images, or conceptual things like colors, subject or a metadata. URIs act as anchor for merging data. In 1997, Metadata vocabularies were defined and used to make statements. A Resource Description Framework (RDF) was prepared. Resource can be anything over Internet identifiable with a URI, Description is the statements about the resource and Framework was a common model for statements using variety of vocabularies. An RDF triple (s, p, o) is defined as "Subject", "Predicate" and "Object" respectively. Conceptually, "p" connects "s" and "o". RDF is a general model for such triples with machine readable formats like RDF/XML.

RDF is useful but is insufficient in solving all possible requirements. To counter the limiting possibilities, a new concept of ontology was introduced. Ontologies define the concepts and relationships for describing and representing an area of knowledge. A new language Web Ontology Language abbreviated as "OWL" was formed. OWL is an extra layer of explanation relying on RDF schema. There are a number of application and web services using RDF schema and OWL to enhance searches as well as cater the need of users. A few examples can be listed as Sun's White Paper and System Handbook collections, Nokia's S60 support portal, Harper's Online Magazine, Oracle's virtual pressroom, Opera's community site, Dow Jones' Synaptica [17].

## 3. THE PROPOSED SYSTEM

Objective is to act as a mediator between human agents who need some service and human or autonomous agents which provide such services. Before we start with the formal description of the model, let us first define few terms that we will use to provide the comprehensive picture.

- *Agent: An entity in the system whether human or autonomous, capable of performing some functionality will be termed as agent. It shows a fundamental unit of behavior based on its perception in the environment. It is intelligent and has a certain level of autonomy to take decisions on its own. It is social so it can interact with other agents of the system.*

- *Activities: The functionalities, these agents can perform will be called as activities. Every agent has some capability based on which it will perform its intended functionality. Based on the capability of an agent, activities can be categorized complex or atomic*

- *Complex activity: A complex activity is one which cannot be performed by a single agent in the environment. Every complex activity must be dissolved in smaller activities.*

- *Atomic activity: If we can find even one agent in the system which is capable enough to carry out complete activity alone, then that task or activity is considered atomic. So it does not depend upon the physical view of a task but ability of an agent to define atomicity of the task.*

- *Environment: Environment in our system is visualized as set of tasks assigned to agents. For example agents involved in the process of request parsing, creating initial semantics, searching and iteratively interacting with domain experts belong to one environment. Then those belonging to negotiation and service provisioning belong to other environment.*

- *Registrar: Registrar is like a central authority responsible for managing agents. It has static knowledge of the agents, their capabilities, domains, location and any other information that is relevant to the services agents provide. It gives a partially centralized view to the architecture controlling limit of distributed functionality.*

- *Domain Expert: Domain experts are the members of a vertical application [18] group defined in semantic web registered with our system with the registrar module. They also function as agents in the system. If a group or organization is registered then a representative agent interacts on their behalf with the system.*

- *Consumer: In this architecture we define a consumer to be a human who needs some kind of help. Consumer submits task request in form of a query. It is assumed that a consumer needs some kind of service from the system. Depending on the service, consumer may be charged by the service provider. Consumer has the liberty to deny any service before the service has already started.*

- *Service Provider: A service provider can be any agent or entity delivering some service to the consumer. Services may be free or may be charged decided by the service provider itself.*

## 3.1 Cognition and rectification of user request

Customer requests do not generally come in the easiest of machine readable formats. Customer request are handled by the communicator agent, which passes it onto the query parser module. Based on the preexisting knowledge base, context reasoning engine, it tries to parse the query for useful information and forms the primitive level semantic of it. An XML file depicting the same is created and sent back to user in a human readable form through communicator agent as depicted in Figure 1.

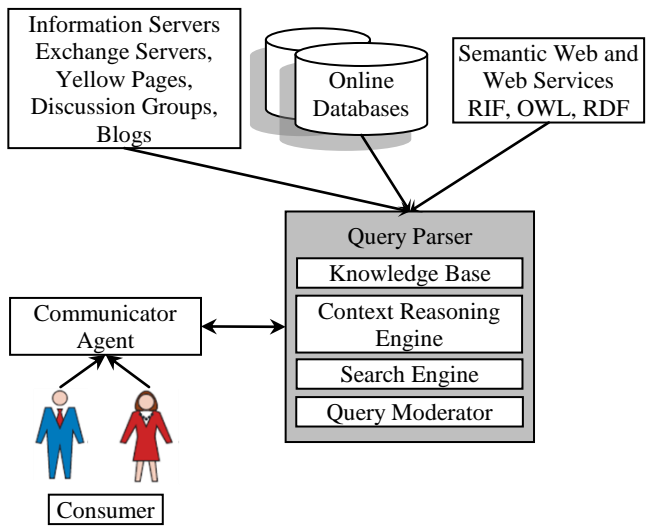

**Figure 1. Accepting consumer request and parsing iteratively for identification of the service requested.**

Sometimes usersthemselvesare not aware of what is needed in a particular problem. User uploads the problem description and hopes to get a solution for it. In such cases, agents cannot rely upon the explanation provided by consumer. What if information is only half truth? System builds an initial query semantic report and sends it to the experts groups of the domain identified in initial report. The initial or primitive understanding is now sent to the domain experts registered in the system through the Registrar module. The domain expert or vertical groups' experts (Figure 2) are manual agents who have better perceptive capability than an autonomous programming agent. Domain experts provide their views on the understanding of problem. They can require more data or different variety of data for better understanding. Sometime they can feel that the problem description does not exactly match their expertise domain so they can refer different agents or domain in the problem description. Collectively this works towards obtaining a better cognition of the problem or request submitted by the customer.

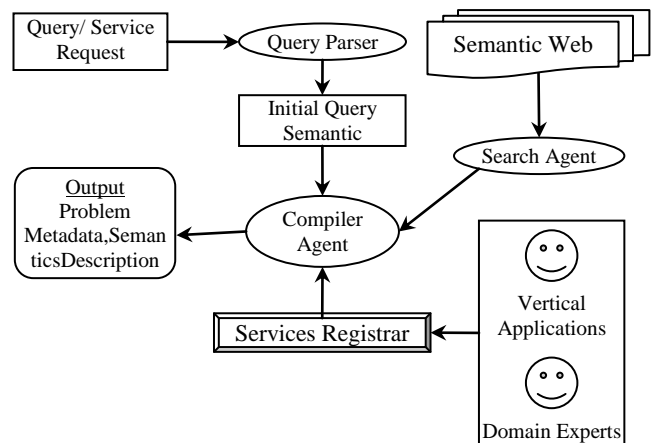

**Figure 2.Entities involved in the identification of a consumer query**

If details provided are not sufficient, consumer can be asked to provide more detail or data. Thus original query is rectified in way both consumer and agent systems obtain a better comprehension. This process is an iterative one which repeats itself as long as either both consumer and domain experts are able to uniquely identify the query or both system and experts decide that there cannot be a solution for the query. In either case, system notifies consumer about the outcome. If it is uniquely identified, system proceeds to the solution development phase, otherwise stops the search.

When agent system develops a mutually agreed upon description of consumer request, it produces a number of descriptor documents. These include Problem Description Document, Vocabulary or Ontology of the request along with any document which is relevant to identify the task uniquely and precisely. These documents help in building inference and developing solution framework.

## 3.2 Developing solution workflow

A workflow is defined as a pattern of activities enabled by a systematic organization of resources, defined roles and mass, energy and information flows, into a work process that can be documented andlearned. Workflows are designed to achieve processing intents of some sort, such as physical transformation, service provision, or information processing. Solution development phase target creating a workflow which when executed provides desired services or solution to the consumer. The workflow is created in such a manner that autonomous agents can follow it and carry out activities.

First step in problem solving is to find the set of experts who have this kind of knowledge. Searching through the web becomes more precise when we have sufficient information and we can target the search space as problem domain. This is again done through Registrar. Note that a separate set of experts are required here as it is not necessary that the expert involved in detecting and explaining the problem can also be expert in actually solving it. For example a pathologist has expertise in interpreting and diagnosing a tumor but lacks surgical experience required to cure it. Nevertheless we do not alienate the idea of an intersection set having agents involved both in identification as well as solution phases. Once experts are found, they are provided complete problem description and their opinion is sought in creating the solution workflow. So system undergoes the similar routine of searching suitable experts as was done in identification phase, provides them the complete problem detail and iteratively works with consumer to fine tune the solution workflow. Customer is always kept in the loop. Automatically creating an optimized workflow for any problem is still a daunting task. So first a tentative workflow is created and presented to experts. Experts supply their opinion and workflow is bettered.

## 3.3 Providing services

A lot of research has been done in multi agent systems to make this part completely free from manual intervention. This section epitomizes the service oriented architecture in its true form. All the participants in this module are autonomous agents, i.e. self interested entities that seek to maximize their private utility [19]. We establish two types of agents here, a consumer (an agent representative of the human consumer) and a provider (again an agent representative of an individual or an organization providing service).

In the Agent Oriented Analysis (AOA), tasks, requirements, services, providers, capabilities and constraints are analyzed from an agent's perspective. Service provisioning consists of following main steps:

1. Workflow description is studied and types of services required are identified. For each service type a list of tasks is created. There may be more than one task possible in any given service type.
2. A list of all possible provider agents is created and mapped to the service type. Each task is considered "Atomic" initially i.e., we assume one agent will be able to finish it to the end. Either agent completes its assigned task or is unable to complete it; there is no concept of partially completed task.
3. If the system could not find an agent having capability to complete a listed task then such a task is considered a "Complex" task and needs to be subdivided in smaller subtasks. This division continues till every task becomes atomic.
4. For each task negotiate with provider agents depending upon the criteria of consumer requirement constraints, and service provider capabilities. Negotiation can be viewed as an interaction mechanism that aims to resolve a conflict of interest between various parties through set rules and protocols [20]. Agents' strategies are very important over here as agents can cooperate or sometime compete to achieve the desired provisioning of services.
   a. Cooperative interactions – agents interact and try to maximize the sum of all their utilities, the social welfare
   b. Competitive interaction – agents interact to satisfy their own personal preferences. Such agents are generally called as selfish or self-interested agents as they try to maximize their own utility function in order to gain greater rewards or optimize the result.

Figure 3 depicts clearly the relation between consumer,

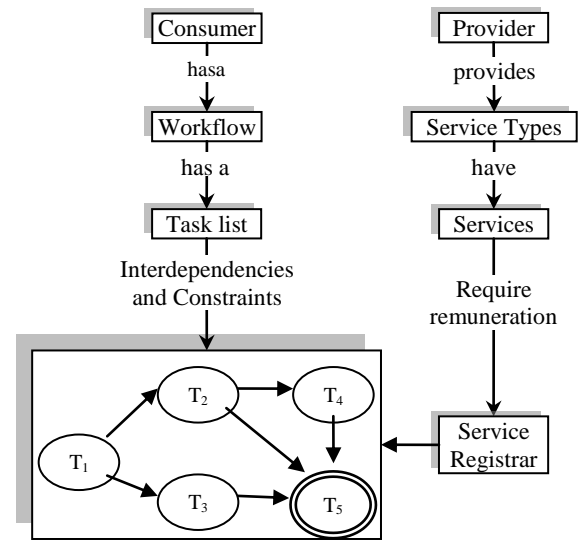

**Figure 3.Relation between workflow of consumer, listed tasks, constraints & interdependencies, services required for these tasks and their service provider**

task, service types, services and provider.

5. If time is relatively a lesser constraint, auction and bidding serves the purpose better as we can draw real life experience from the examples of tender invited by an organization.

## 3.4 Design of the service model

As defined earlier in Section 3, any entity performing a unit of work is treated as an *agent* in our model. Every agent has a certain set of defined capabilities. Agents act perceiving the change in its environment. When it has certain task to complete, it's the agent's responsibility not only to complete the task but also do it in the best possible of ways.The actions, an agent undertakes to fulfill its responsibilities is captured by the idea of a utility function. Utility functions are useful in describing the tradeoff every agent must make. For example, agent receives payment for searching and delivering information and incurs cost of the electricity used as well as the benefits it missed had it been searching and delivering information to some other consumer. If we translate all these payments andcosts into utility function then we can easily study the tradeoffs among them [21]. Agents interact by sending and receiving messages. When an agent initiates interaction by sending a message, it becomes requester agent and the agent it sends message to become a provider agent.

*Messages* are sent to communicate between agents delivering web services. When an agent wants to avail some service, it sends request message to the agent representing this particular type of service with input requirement semantics. Provider agent acknowledges the request only if it is interested in fulfilling it and explains the terms and conditions of providing services. For example high priority programs which execute immediately cost higher than those put into batch processing mode. Thus when requester agent and provider agent become known to each other, they understand the input semantics and Web Service Description (WSD) formally, service is provided. Structure of the message is also defined in WSD. A message can be a simple HTTP GET request or it can be a SOAP XML. In our model, every service provider is registered to the Registrar. Initial communication of knowing the service provider and requester occur through the Registrar, but once their representative agents acknowledge each other, they can negotiate freely without intervention of Registrar. In our system messages are sent through using Simple Object Access Protocol (SOAP) exchanging XML documents over the Internet, using Hypertext TransferProtocol (HTTP) [22].

*Web Services* registration and discovery is done using Universal Description, Discovery and Integration (UDDI) directory service. Businesses use Web Services Description Language (WSDL) to describe relevant information to potential consumers and advertise services. In our model any activity that results in some fruitful physical outcome is termed as service. Starting from the very first interaction of system with consumer through communicator agent to parsing request, understanding semantics of the request, consulting domain experts, identification of the problem, searching and consulting solution experts, creating solution workflow, finding service providers, negotiating, contracting and getting services completed, all include service definition. Every expert involved, whether a human agent giving feedback or suggestion or an autonomous agent searching the semantic web for some information provides service. Some services are free for example accessing freely available information on semantic web, finding suitable service providers through WSDL description provided in the directory service, contacting providers through SOAP messages and negotiating a binding contract etc. Only those services will be charged which has some copyright to it and/or provider requires some remuneration. When we define a service in our system, it should be clear that all services are provided through the Registrar module, which acts as first level of central authority over collaborative autonomous agents acting towards achieving solution.

## 4. ANALYSIS OF THE MODEL

We started building the system with the most essential of requirements expected from a multi agent system delivering service. Initial requirements can be classified as *functional requirements* and *quality requirements*.

### 4.1 Functional requirements

- *Autonomy:* Agents are autonomous as long as it a utility function to maximize. There are dominant agents whose task is just to monitor the progress of task. In case an agent fails to produce output, dominant agents reassign the task to some other agent. Though dominant agents do not interfere in the functionality of an individual agent they do put some kind of organization. Agents are free to enter the environment and leave it as required.

- *Social interaction:* As per the inherent definition an agent here is able to communicate with other agents through message passing. These messages are XML files describing requirements, requesting services or reporting status. Though we have put some restriction in interaction here that agents of one environment would not be able to interact with different environment agents.

- *Collaboration:* Agents must collaborate to sort out an assigned task in order to maximize their utility functions. Collaboration can take cooperative form if it maximizes the generic output else it can also take a competitive form when being selfish results in an optimized result, e.g. in the case of auction, bidding agents compete with each other resulting in a better deal for the consumer.

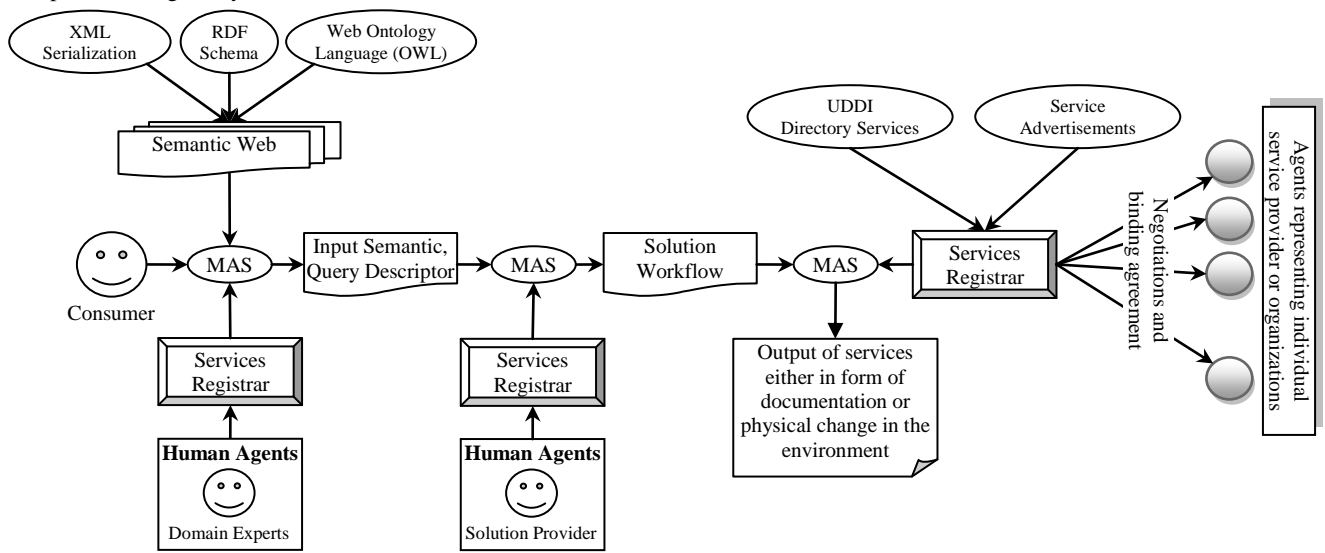

**Figure 4. Comprehensive view of the service depicting life stages of consumer request including query cognition resulting in Input semantics, then reconstruction as a solution workflow, and finally resolution in the form of services provided**

- *Distributed functionality:* Agents act in a distributed manner when tasks are temporally independent. Services like communication, data transfer, searching, negotiating, management etc. take place without intervention of any centralized processing unit.

- *Reusability:* We have used several off-the shelf components in the overall development recognizing reusability of agent programs.

## 4.2 Quality requirements

- *Ease of use:* User interface should be easily understood and easy to use. User can upload query description in the form of a text file. Status updates are provided back to user and provisions are made to constantly involve user in the loop.

- *Modifiability:* Agents can be modified without the need to change whole architecture. As long as interaction mechanism is coherent, internal functionality is a black box for the environment.

- *Reliability:* As long as there is no network failure, agents are reliable. Agents are able to communicate as soon as network re-establishes.

- *Performance:* Although requirement criterion specifies the system to accomplish tasks in a reasonable amount of time with optimized results but there are so many smaller modules with open research problem interoperating that no claim can be made at this stage.

- *Security:* Security is managed through the Registrar module of system as only registered entities can use the system.

Figure 4, depicts the overall architecture with stages of the life cycle of a consumer request starting from when request is made to the point when services in terms of satisfactory solution are provided back to user. Until a binding contract is made, consumer is free to abandon request anytime. Provisions must be made to ensure through the stages that the consumer is still interested in find a solution for her/her query. All the activities should be abandoned as soon as customer backs off.

## 5. CONCLUSION AND FUTURE CHALLANGES

In this paper we describe the architecture combining efficiencies of Multi Agent System with Semantic Web to provide better query understanding, search, and resolution in a modular approach. We cannot claim to have an optimized solution as there are several open research problems in the system. We provide software engineering platform for efficient merger of coherent technological domains. Multi agent systems research is a very hot field in terms of providing different kind of services in extremely coordinated fashion as shown in disaster management domain by the recently concluded Aladdin project [23]. Problems such as automatic creation of efficient solution workflow, negotiating and legally binding service providers, remuneration management, workflow mitigation, security, data integration, building commitment, trust, and belief model are some of the areas where rapid research is going on. We have set a target in our architecture focusing smaller modules and trying to refine the functionalities as well as standardize the agent communication on semantic web.